\documentclass[prd,eqsecnum,preprint,showpacs]{revtex4}

\usepackage{stmaryrd}
\usepackage{amsmath}
\usepackage{amssymb}
\usepackage{graphicx}
\usepackage{textcomp}
\usepackage{calrsfs}
\usepackage{yfonts}

\newcommand{\be}{\begin{equation}}
\newcommand{\ee}{\end{equation}}
\newcommand{\ben}{\begin{equation*}}
\newcommand{\een}{\end{equation*}}
\newcommand{\bea}{\begin{eqnarray}}
\newcommand{\eea}{\end{eqnarray}}

\renewcommand{\Im}{\mbox{Im}}
\renewcommand{\Re}{\mbox{Re}}
\newcommand{\re}{\mbox{Re}}
\newcommand{\im}{\mbox{Im}}

\newcommand{\dif}{\text{d}}

\DeclareMathOperator{\Tr}{Tr}
\DeclareMathOperator{\tr}{tr}
\DeclareMathOperator{\Ai}{Ai}
\DeclareMathOperator{\Bi}{Bi}
\DeclareMathOperator{\arcsec}{arcsec}

\begin{document}

\title{Casimir Effect for a Semitransparent Wedge and an Annular Piston}

\date{\today}

\author{Kimball A. Milton}\email{milton@nhn.ou.edu}
\author{Jef Wagner}\email{wagner@nhn.ou.edu}
\affiliation{Oklahoma Center for High Energy Physics and H.L. Dodge
 Department of
Physics and Astronomy, University of Oklahoma, Norman, OK 73019-2061, USA}
\author{Klaus Kirsten}\email{Klaus_Kirsten@baylor.edu}
\affiliation{Department of Mathematics, Baylor University,
Waco, TX 76798-7328, USA}

\begin{abstract}
  We consider the Casimir energy due to a massless scalar field
in a geometry of an infinite
wedge closed by a Dirichlet circular cylinder, where the
wedge is formed by $\delta$-function potentials, so-called semitransparent
boundaries.  A finite expression for the Casimir energy
corresponding to the arc and the presence of both semitransparent
potentials is obtained, from which the torque on the sidewalls can be
derived. The most interesting part of the calculation
is the nontrivial nature of the angular mode functions.  Numerical results
are obtained which are closely analogous to those recently found for
a magnetodielectric wedge, with the same speed of light on both sides
of the wedge boundaries.  Alternative methods are developed for annular
regions with radial semitransparent potentials,
 based on reduced Green's functions for the angular dependence,
which allows calculations using the multiple-scattering formalism.
Numerical results corresponding to the torque on the radial plates
are likewise computed, which generalize those for the wedge geometry.
Generally useful formulas for calculating Casimir energies in separable
geometries are derived.
\end{abstract}

\maketitle
\section{Introduction}
The Casimir effect \cite{casimir48}, which was originally conceived
as the attraction between parallel perfectly conducting plates, may
be regarded as due to the fluctuations of the electromagnetic field
in  the quantum vacuum. In the past six decades, this phenomenon has
been generalized to many different types of fields and to a variety
of geometries and topologies.
Recent reviews of the Casimir effect include
Refs.~\cite{BookMilton01, milton04, lamoreaux05,buhmann07,bordagbook}.

In this paper we will illustrate some new features that arise, for example,
in cylindrical geometries in which the cylindrical symmetry is broken.  In
the past three decades there have been many works on problems possessing
cylindrical symmetry, starting with the calculation of  the Casimir energy
of an infinitely long perfectly conducting
cylindrical shell \cite{deraad81}. The more physical but also
significantly more involved case of a dielectric cylinder was considered
more recently \cite{brevik94, gosdzinsky98,milton99, lambiase99,
caveroPelaez05, romeo05,brevik07}.  Particularly germane to the present
work is the calculation of the Casimir effect for a scalar field interior
and exterior to a cylindrical $\delta$-function potential,
a so-called semitransparent cylinder \cite{caveroPelaez06};
in the weak-coupling limit, both the semitransparent cylinder and the
dielectric cylinder have vanishing Casimir energy.

The infinite wedge is closely related to the cylindrical geometry.
This problem
was first considered in the late seventies \cite{dowker78, deutsch79} as part
of the still ongoing debate about how to interpret various divergences in
quantum field theory with sharp boundaries and whether self-energies
of objects have any physical significance. Since then, variations on
this idea of the
wedge have been treated by several authors \cite{brevik96,brevik98,
brevik01,nesterenko02, razmi05}, and reviewed in
Ref.~\cite{BookMostepanenko97}. A wedge with a coaxial cylindrical shell was
considered by Nesterenko et al.\ \cite{nesterenko01,nesterenko03},
and the corresponding local stresses were investigated by Saharian and
collatorators \cite{rezaeian02,saharian07, saharian09,saharian05,saharian08}.
The interaction of an atom with a wedge was studied in Refs.~%
\cite{barton87,skipsey05, skipsey06,mendes08, rosa08};
this geometry is that of the experiment by Sukenik et al.\ carried out
more than 15 years ago \cite{sukenik93}. Recently,
Brevik et al.~\cite{brevik09}  calculated the Casimir energy of a
magnetodielectric cylinder intercut by a perfectly reflecting wedge filled
with magnetodielectric material.
In all of these studies  the assumption was made
that the wedge be bounded by perfectly conducting walls.

Although wedges defined by perfect conductors or Dirichlet boundaries
break cylindrical symmetry, they do so in an easily understood way.
When cylindrical symmetry is present, the azimuthal quantum number $\nu$ ranges
from $-\infty$ to $\infty$ by integer steps.  With a perfect conductor,
which forces the tangential electric field to vanish on the surface,
$\nu$ takes on values which are related to the opening angle $\alpha$
of the wedge, $\nu=\pi m/\alpha$, where $m$ is an integer.  But what if
the wedge boundaries are not perfect?  Recently, Ellingsen et
al.~\cite{ellingsen09} considered just such a case, where the wedge was
defined by the interface between two magnetodielectric media, where
the interior sector of the wedge had permittivity $\varepsilon_1$ and
permeability $\mu_1$, while the exterior sector had permittivity
$\varepsilon_2$ and permeability $\mu_2$.  The geometry was completed
by inserting a perfectly conducting circular cylinder
of radius $a$ centered on the wedge axis. To assure a finite result,
as well as separability of the problem,
the further assumption was made there that the speed of light in both
media was the same: $\varepsilon_1\mu_1=\varepsilon_2\mu_2.$  In this
case the azimuthal quantum number had to be determined by a transcendental
equation, which was implemented in the calculation through use of the
so-called argument principle \cite{vankampen68}, which is just the residue
theorem.

\begin{figure}[tb]
  \begin{center}
  \includegraphics[width=3in]{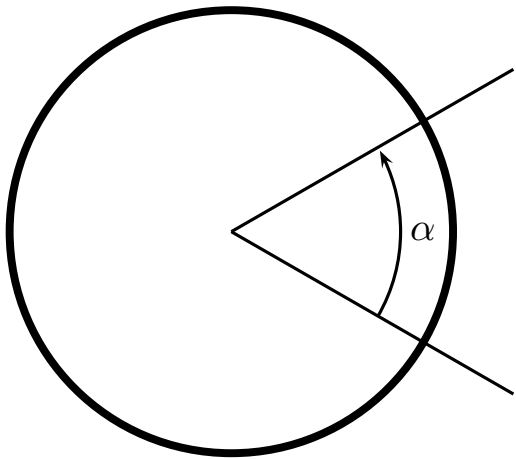}
  \caption{A Dirichlet cylinder intersecting with a coaxial wedge made of
semitransparent plates.}
  \label{fig1}
  \end{center}
\end{figure}

In this paper, we further illustrate this nontrivial azimuthal dependence
by considering a similar wedge geometry, in which the infinite wedge is formed
by two planar $\delta$-function potentials, making a dihedral angle
$\alpha\in [0,\pi]$, closed by a coaxial Dirichlet circular cylinder of
radius $a$.  See Fig.~\ref{fig1}.
We calculate the Casimir energy of a massless scalar field subject to these
boundary conditions.  Because that energy is divergent, we compute the
energy relative to that when the radius of the cylinder is infinite,
and when neither or only one of the wedge boundaries is present.  Thus, we are
computing the energy of mutual interaction between the three boundaries.
The results, which are rather easily found numerically, are very similar
to those found for the electromagnetic field in a perfectly conducting
cylinder with a magnetodielectric wedge, as considered in
Ref.~\cite{ellingsen09}.  We describe the geometry in terms of cylindrical
coordinates, $(\rho,\theta,z)$, with the origin lying along the
cylinder axis.

Because the interest in this problem largely lies in the angular dependence,
it is natural to approach the problem in an unconventional way, in which
the reduced Green's function refers to the azimuthal, not the radial
coordinate.  Technically, that approach requires consideration of an annular
region, which we describe in Sec.~\ref{sec:alt}.  This approach
also allows use of the multiple-scattering technique,
and should have application
to more complicated geometries, such as the interaction between
hyperboloids.  We can think of the radial planes between the
concentric cylinders as forming an annular piston, and we have
computed numerically the Casimir attractive torque between those planes.
An alternative approach to the determination of the Casimir
energy for any such angular potential is described in Sec.~\ref{sec:theta}.
The radial functions encountered in these wedge problems are modified
Bessel functions of imaginary order; since these are rather infrequently
described in the literature, we collect some relevant properties in
Appendix \ref{app:b}.  Required integrals over the squared radial
functions may be evaluated using identities described in Appendix \ref{sec4}.

%%%%%%%%%%%%%%%%%%%%%%%%%%%%%%%%%%%%%%%%%%%%%%%%%%
%%%%%%%%%%%%%%%%%%%%%%%%%%%%%%%%%%%%%%%%%%%%%%%%%%
%%%%%%%%%%%%%%%%%%%%%%%%%%%%%%%%%%%%%%%%%%%%%%%%%%

\section{Semitransparent wedge}\label{sec-st}
In this paper we consider a massless scalar model, in which the
wedge is described by a $\delta$-function potential, $V(\rho,\theta)
=v(\theta)/\rho^2$,
\be
v(\theta)=\lambda_1\delta(\theta-\alpha/2)+\lambda_2\delta(\theta+\alpha/2).
\label{dfpot}
\ee
This has the diaphanous property of preserving the speed of
light both within and outside the wedge.
This wedge is superimposed on a coaxial circular cylindrical shell, of radius
$a$, on which the scalar field $\phi$ vanishes.  To calculate the
Casimir energy, we can use the formula \cite{BookMilton01}
\be
E=\frac1{2i}\int_{-\infty}^\infty \frac{d\omega}{2\pi}\int (d\mathbf{r})\,
2\omega^2\mathcal{G}(\mathbf{r,r};\omega),\label{magic}
\ee
where $\mathcal{G}$ is the Green's function for the situation under
consideration, satisfying
\be
\left(-\nabla^2+V(\rho,\theta)-\omega^2\right)\mathcal{G}(\mathbf{r,r'};
\omega)=\delta(\mathbf{r-r'}).
\ee

 We can solve this cylindrical
problem in terms of the two-dimensional Green's function $G$,
\be
\mathcal{G}(\mathbf{r,r'};\omega)=\int_{-\infty}^\infty \frac{dk_z}{2\pi}
e^{ik_z(z-z')}G(\rho,\theta;\rho',\theta'),
\ee
 which satisfies
\bea
\left[-\frac1\rho\frac\partial{\partial \rho}\rho
\frac\partial{\partial\rho}
+\kappa^2-\frac1{\rho^2}\frac{\partial^2}{\partial\theta^2}
+\frac{v(\theta)}{\rho^2}
\right]G(\rho,\theta;\rho',\theta')=\frac1\rho\delta(\rho-\rho')
\delta(\theta-\theta'),\label{twogr}
\eea
where $\kappa^2=k_z^2-\omega^2$.
This separates into two equations, one for the angular eigenfunction
$\Theta_\nu(\theta)$,
\be
\left[-\frac{\partial^2}{\partial\theta^2}+v(\theta)\right]\Theta_\nu(\theta)
=\nu^2\Theta_\nu(\theta),\label{eveq}
\ee
where we have assumed that the azimuthal eigenfunctions are normalized
according to
\be
\int_{-\pi}^\pi d\theta\, \Theta_\nu(\theta)\Theta^*_{\nu'}(\theta)=\delta_{\nu
\nu'};
\ee
orthogonality of the eigenfunctions follows from the Sturm-Liouville nature
of the problem.
Now the two-dimensional Green's function can be constructed as
\be
G(\rho,\theta;\rho',\theta')=\sum_\nu
\Theta_\nu(\theta)\Theta_\nu^*(\theta')g_\nu(\rho,\rho').\label{gfconst1}
\ee
The radial reduced Green's function satisfies
\be
\left[-\frac1\rho\frac\partial{\partial \rho}\rho\frac\partial{\partial \rho}
+\kappa^2
+\frac{\nu^2}{\rho^2}\right]g_\nu(\rho,\rho')=\frac1\rho\delta(\rho-\rho').\label{redgre}
\ee
The latter, for a Dirichlet circle at $\rho=a$, has the familiar solution,
\begin{subequations}
\bea
g_\nu(\rho,\rho')&=&I_\nu(\kappa \rho_<)K_\nu(\kappa \rho_>)-I_\nu(\kappa \rho)
I_\nu(\kappa \rho')
\frac{K_\nu(\kappa a)}{I_\nu(\kappa a)},\quad \rho,\rho'<a,\\
g_\nu(\rho,\rho')&=&I_\nu(\kappa \rho_<)K_\nu(\kappa \rho_>)-K_\nu(\kappa \rho)
K_\nu(\kappa \rho')
\frac{I_\nu(\kappa a)}{K_\nu(\kappa a)},\quad \rho,\rho'>a.\label{gradcir}
\eea
\end{subequations}

The azimuthal eigenvalue $\nu$ is determined by Eq.~(\ref{eveq}).  For the
wedge $\delta$-function potential (\ref{dfpot}) it is easy to determine
$\nu$ by writing the solutions to Eq.~(\ref{eveq})
as linear combinations of $e^{\pm i\nu\theta}$,
with different coefficients in the sectors $|\theta|<\alpha/2$ and $\pi\ge
|\theta|>\alpha/2$. Continuity of the function, and discontinuity of
its derivative, are imposed at the wedge boundaries. The four simultaneous
linear homogeneous equations have a solution only if the secular equation
is satisfied:
\be
0=D(\nu)
=\sin^2\nu(\alpha-\pi)-\left(1-\frac{4\nu^2}{\lambda_1\lambda_2}\right)
\sin^2\pi\nu
-\left(\frac{\nu}{\lambda_1}+\frac\nu{\lambda_2}\right)
\sin2\pi\nu.\label{a6}
\ee
Because we recognize that the reflection coefficient for a single
$\delta$-function interface is
$r_i=(1+2i\nu/\lambda_i)^{-1}$, implying that
\be
\re \left(r_1^{-1}r_2^{-1}\right)=1-\frac{4\nu^2}{\lambda_1\lambda_2},\quad
\im \left(r_1^{-1}r_2^{-1}\right)
=\frac{2\nu}{\lambda_1}+\frac{2\nu}{\lambda_2},\label{a7}
\ee
we see that this dispersion relation coincides with that
found in Ref.~\cite{ellingsen09} when the reflection coefficient is purely
real.
Note that the $\nu=0$ root of Eq.~(\ref{a6}) is spurious and must be excluded;
unlike for the magnetodielectric wedge,
there are no $\nu=0$ modes for the semitransparent wedge.

Now using the general formula (\ref{magic}) we compute
the Casimir energy per length from
\be
\mathcal{E}=\frac1{2i}\int_{-\infty}^\infty \frac{d\omega}{2\pi}2\omega^2
\int_{-\infty}^\infty \frac{dk}{2\pi}\sum_\nu \int_0^\infty d\rho\,\rho\,
g_\nu(\rho,\rho).
\ee
Note that we do not need to know the eigenfunctions $\Theta_\nu$,
only the eigenvalues $\nu$.

\begin{figure}[tb]
  \begin{center}
  \includegraphics[width=3in]{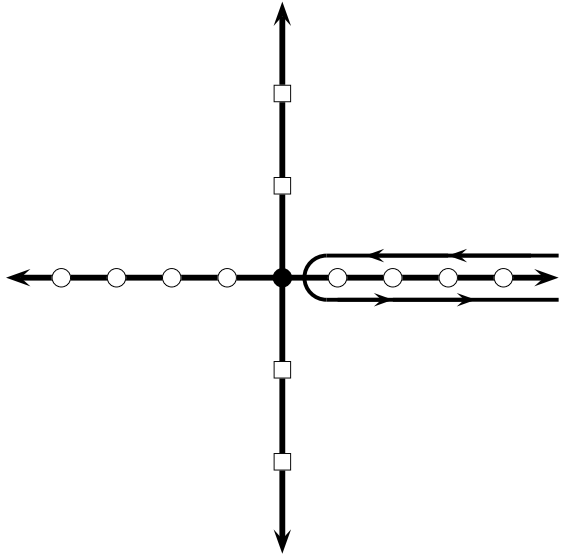}
  \caption{Contour of integration $\gamma$ for the argument principle
(\ref{arg-prin}). Shown also are singularities of the integrand
(\ref{eform}) along the real and imaginary $\nu$ axes.}
  \label{fig-gamma}
  \end{center}
\end{figure}

The apparent difficulty, that the eigenvalue condition for $\nu$ cannot
be explicitly solved, may be resolved through
enforcing the eigenvalue condition by the argument principle
\cite{vankampen68, bordag96,
BookParsegian06,brevik01b,nesterenko06,nesterenko08},
 which gives a sum over non-explicit
eigenvalues in terms of a contour integral around the real line,
\begin{equation}
  \sum_\nu f(\nu)=\frac{1}{2\pi i}\int\limits_\gamma d\nu
  \left(\frac{d}{d\nu}\ln D(\nu)\right) f(\nu).
\label{arg-prin}
\end{equation}
The contour of integration $\gamma$ is illustrated in Fig.~\ref{fig-gamma}.
Thus, we have the expression after making the Euclidean rotation,
$\omega\to i\zeta$, and converting to polar coordinates,
\be
\zeta=\kappa \cos\varphi,\quad k=\kappa\sin\varphi,
\ee
\be
\mathcal{E}=-\frac1{8\pi^2i}\int_0^\infty d\kappa\,\kappa^3\int_\gamma
 d\nu\left(\frac{d}{d\nu}\ln D(\nu)\right)\int_0^\infty d\rho\,
\rho \, g_{\nu}(\rho,\rho).\label{eform}
\ee

This formal expression is rather evidently divergent.  We are seeking the
mutual interaction energy due to the three boundaries, the two sides of the
wedge and the circular arc.  Therefore, we first must
subtract off the free radial Green's function without the circle at $\rho=a$,
which then implies
\be
\int_0^\infty d\rho\,\rho\,g_{\nu}(\rho,\rho)\to\frac{a}{2\kappa}
\frac{d}{d\kappa a}
\ln[I_{\nu}(\kappa a)K_{\nu}(\kappa a)].\label{intofg}
\ee
(The familiar form of this expression is quite general, as illustrated
in Appendix \ref{sec4}.)
We further want to
remove the term present without the wedge potential:
\be
D(\nu)\to\tilde D(\nu)=\frac{\lambda_1\lambda_2}{4\nu^2}
\frac{D(\nu)}{\sin^2\pi\nu}.
\ee
The resulting energy is still not finite.  The reason is that it contains
the self-energy
of a single $\delta$-function potential crossed by the circular cylinder.

Therefore, we still must remove that part of $\tilde D$ due to a single
potential, which may be obtained by setting $\lambda_2$ (or $\lambda_1$)
equal to zero:
\be
\tilde D_1(\nu)=1-\frac{\lambda_1}{2\nu}\cot \nu\pi,\label{d-1pot}
\ee
so the final form of the dispersion function is obtained by replacing
\bea
\tilde D(\nu)\to\hat D(\nu)&=&\frac{\tilde D(\nu)}{\tilde D_1(\nu)
\tilde D_2(\nu)}\nonumber\\
&=&\frac{\lambda_1\lambda_2\sin^2\nu(\alpha-\pi)/
\sin^2\nu\pi+4\nu^2-\lambda_1\lambda_2-2\nu(\lambda_1+\lambda_2)
\cot\nu\pi}{(2\nu-\lambda_1\cot\nu\pi)(2\nu-\lambda_2\cot\nu\pi)}.
\nonumber\\ \label{fulldee}
\eea
(Although the spurious $\nu=0$ root is still present, it may be checked
that this gives rise to an irrelevant divergent constant in the energy.)

\begin{figure}[tb]
  \begin{center}
  \includegraphics[width=3in]{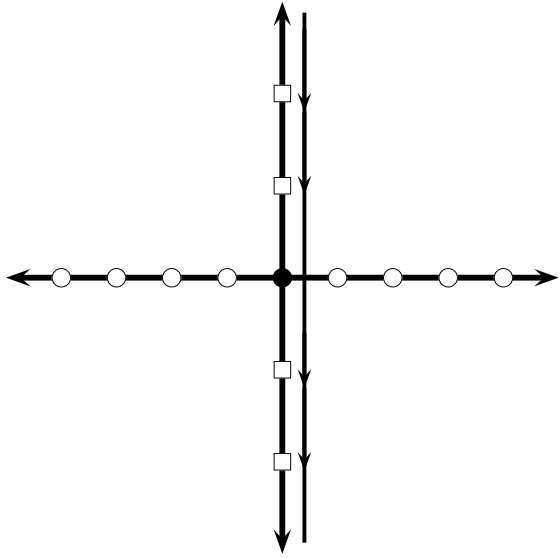}
  \caption{Contour of integration for the $\eta$ integral in
Eq.~(\ref{expression}).}
  \label{fig-unfold}
  \end{center}
\end{figure}

It is now easy to see that the integrand in the expression for the energy
falls off exponentially fast for large $\nu$ in the right-half complex
$\nu$ plane, except along the real $\nu$ axis, where an
exponential convergence factor may be inserted.  
In particular, for $\nu=i\eta$,
 $\eta\gg1$, $\hat D(i\eta)$  differs only exponentially from unity:
\be
\hat D(i\eta)\sim 1-\frac{\lambda_1\lambda_2}{(2\eta+\lambda_1)
(2\eta+\lambda_2)}e^{-2\eta \alpha}.\label{D-asym}
\ee
Then it is permissible to unfold $\gamma$ and convert the contour
to one running parallel to the imaginary axis as shown in
Fig.~\ref{fig-unfold}.
For imaginary $\nu$, $\nu=i\eta$, the dispersion functions become
\be
\tilde D_1(i\eta)=1+\frac{\lambda_1}{2\eta}\coth \eta\pi,\label{d-1pot-e}
\ee
and
\be
\hat D(i\eta)=\frac{-\lambda_1\lambda_2\sinh^2\eta(\alpha-\pi)/
\sinh^2\eta\pi+4\eta^2+\lambda_1\lambda_2+2\eta(\lambda_1+\lambda_2)
\coth\eta\pi}{(2\eta+\lambda_1\coth\eta\pi)(2\eta+\lambda_2\coth\eta\pi)}.
\label{fulldee-e}
\ee

Because of Eq.~(\ref{D-asym}),
the resulting expression  for the Casimir energy is manifestly convergent.
This can be further simplified by noting that
$\frac{d}{d\eta}\ln \hat D(i\eta)$ is odd, which
eliminates the $K_\nu$ in Eq.~(\ref{intofg}), and then yields the
expression
\be
\mathcal{E}=\frac1{8\pi^2 a^2}\int_0^\infty dx\,x^2\int_0^\infty d\eta
\left(\frac{d}{d\eta}\ln\hat D(i\eta)\right) \frac{d}{dx}
\arctan\frac{K_{i\eta}(x)}{L_{i\eta}(x)},\label{expression}
\ee
in terms of
\begin{subequations}
\label{def-k-l}
\bea
K_\mu(x)&=&-\frac\pi{2\sin\pi\mu}\left[I_\mu(x)-I_{-\mu}(x)\right],\\
L_\mu(x)&=&\frac{i\pi}{2\sin\pi\mu}\left[I_\mu(x)+I_{-\mu}(x)\right],
\eea
\end{subequations}
where both $L_{i\eta}(x)$ and $K_{i\eta}(x)$ are real for real $\eta$ and
$x$, and
\be
I_{i\eta}(x)=\frac{\sinh\eta\pi}\pi[L_{i\eta}(x)-iK_{i\eta}(x)].
\ee
We should further note that the arctangent appearing in Eq.~(\ref{expression})
is not the principal value, but rather the smooth function in which the
phase is accumulated.  (Some properties of the modified Bessel functions
of imaginary order are collected in Appendix \ref{app:b}.)

Now we turn to the numerical evaluation of this expression.

\section{Numerical Evaluation of Casimir Energy for the
Semitransparent Wedge}

\begin{figure}[tb]
  \begin{center}
  \includegraphics[width=3in]{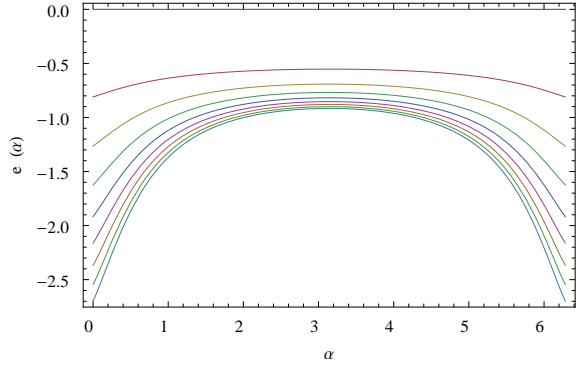}
  \caption{Casimir energy for a semitransparent wedge embedded in a
Dirichlet cylinder, as a function of the dihedral angle $\alpha$.
Shown in order from highest to lowest are the energies (\ref{energy}) for
$\lambda_1=\lambda_2=0.5$ to 4.0, by steps of 0.5.}
  \label{fig2}
  \end{center}
\end{figure}

\begin{figure}[tb]
  \begin{center}
  \includegraphics[width=3in]{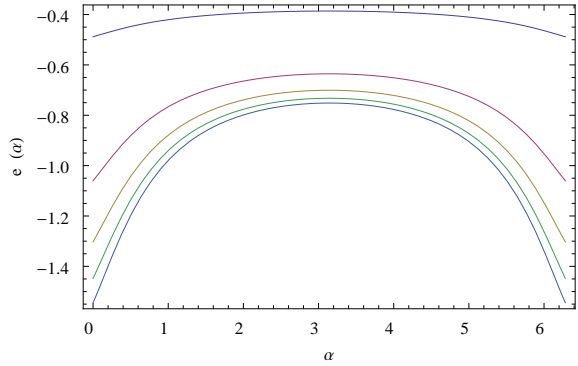}
  \caption{Casimir energy for a semitransparent wedge embedded in a
Dirichlet cylinder, as a function of the dihedral angle $\alpha$.
Shown in order from highest to lowest are the energies (\ref{energy}) for
$\lambda_1=1$ and $\lambda_2=0.1$ to 2.1, by steps of 0.5.}
  \label{fig3}
  \end{center}
\end{figure}

It is actually quite easy to evaluate Eq.~(\ref{expression}), because
the $\frac{d}{d\eta}\log\hat D$ function is strongly peaked for small
$\eta$, except for extremely small values of the dihedral angle $\alpha$.
The difficulty numerically is that $K_{i\eta}(x)/L_{i\eta}(x)$ is an
extremely oscillatory function of $x$ for $x<\eta$, becoming infinitely
oscillatory as $x\to 0$.  For $x>\eta$, the ratio of modified Bessel
functions of imaginary order monotonically and exponentially approaches
zero.  (For incomplete asymptotic information about Bessel functions
of imaginary order see Refs.~\cite{dunster90,olver}; see also
Appendix \ref{app:b}.)
The function
\be
h(\eta)=\int_0^\infty dx\,x^2\frac{d}{dx}\arctan\frac{K_{i\eta}(x)}{L_{i\eta}
(x)},
\ee
however, is very smooth. (It vanishes at $\eta=0$, so the spurious zero
mode should not contribute.)
To evaluate the double integral, we compute
$h$ at a finite number of discrete points, form a spline approximation
which is indistinguishable from $h$, and then evaluate the function
\be
e(\alpha)=\int_0^\infty d\eta\,h(\eta)\frac{d}{d\eta}\ln\hat D(i\eta),
\ee
numerically. (This strategy is similar to that employed in
Ref.~\cite{ellingsen09}.)
The integrand here is quite strongly peaked in a neighborhood
of the origin. The Casimir energy, with the indicated
subtractions, is
\be
\mathcal{E}=\frac1{8\pi^2 a^2}e(\alpha).\label{energy}
\ee
 The results found by this strategy are
shown in Figs.~\ref{fig2} and \ref{fig3}.

These graphs are very reminiscent of those found in Ref.~\cite{ellingsen09}
for the magnetodielectric wedge.  In particular, we note that the energies
are finite for all $\alpha$, but as $\lambda\to\infty$, the limit
corresponding to a Dirichlet boundary, the energy diverges as $\alpha\to 0$
or $2\pi$; the same phenomena was observed in Ref.~\cite{ellingsen09} for
the perfectly conducting wedge limit, treated previously in
Ref.~\cite{brevik09}.  This energy should be observable as a torque on the two
semitransparent plates, $\tau(\alpha)=-\frac\partial{\partial\alpha}\mathcal{E}
(\alpha)$, which is, as expected, attractive.  (The divergence associated
with the apex of the wedge has been subtracted.)

\section{Alternative
calculation of Casimir energy for semitransparent wedge in an annulus}
\label{sec:alt}
\begin{figure}[tb]
  \begin{center}
    \includegraphics[width=1.7in]{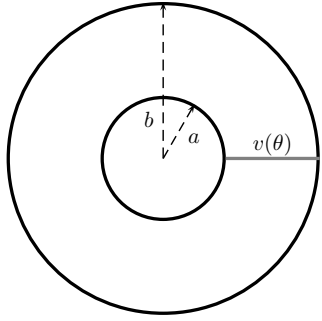}
    \caption{\label{fig-ap}The annular geometry considered.}
  \end{center}
\end{figure}

We start from the formula (\ref{magic}) for the Casimir energy in terms of the
Green's function,
\begin{equation}\label{cas_en}
  E=\frac{1}{2i}\int\frac{d \omega}{2\pi}2\omega^2
  \Tr (\mathcal{G}-\mathcal{G}^{(0)}),
\end{equation}
where the trace denotes the integration over spatial coordinates,
and we have again subtracted out the vacuum contribution.
The Green's function $\mathcal{G}(\mathbf{r,r'})$ will satisfy the equation
\begin{equation}
  \left[-\nabla^2-\omega^2+V(\mathbf{r})\right]\mathcal{G}(\mathbf{r,r'})=
  \delta(\mathbf{r-r'}),
\end{equation}
while the free Green's function $\mathcal{G}^{(0)}$
satisfies the same equation with
$V(\mathbf{r})=0$.  Once again we specialize to the cylindrical
geometry, but now defined in an annulus.  Specifically, we require that
the boundary conditions on the Green's function are
that it vanishes at $\rho=a$ and $\rho=b$ with $b>a$, that is, it
satisfies Dirichlet boundary conditions on two concentric circles. (We
will see the necessity for both an inner and an outer boundary in the
following.) If the potential has
the form $V(\mathbf{r}) = v(\theta)/\rho^2$ then we can use separation of
variables to write the Green's function as, in terms of the separation
constant $\eta$,
\begin{equation}\label{expanded_g}
  \mathcal{G}(\mathbf{r,r'};\omega)=\int_{-\infty}^\infty
  \frac{d k_z}{2\pi}e^{i k_z(z-z')}
  \sum_\eta R_\eta(\rho;\omega,k_z)R_\eta(\rho';\omega,k_z)
  g_\eta(\theta,\theta').
\end{equation}
The geometry we are considering is illustrated in Fig.~\ref{fig-ap}.
Note that instead of expanding in eigenfunctions of $\theta$ as in
Eq.~(\ref{gfconst1}), we have expanded in terms of radial eigenfunctions.
(This cannot be done without the inner boundary---that is, this alternative
separation works for an annulus but not for a disk.)
The $R$ functions are normalized radial eigenfunctions of the eigenvalue
problem,
\begin{equation}\label{rad_de}
  \left[-\rho \frac{d}{d\rho} \rho \frac{d}{d\rho} -(\omega^2 -k_z^2)\rho^2
\right]  R_\eta(\rho;\omega,k_z)=\eta^2R_\eta(\rho;\omega,k_z),
\end{equation}
with boundary values $R_\eta(a;\omega,k_z)=R_\eta(b;\omega,k_z)=0$. The
$g_\eta$ is the reduced Green's function in the azimuthal coordinates that
satisfies the equation
\begin{equation}
  \left[-\frac{d^2}{d\theta^2} +\eta^2+v(\theta)\right]g_\eta(\theta,\theta')=
  \delta(\theta-\theta'),
\end{equation}
with periodic boundary conditions. Finally inserting Eq.~(\ref{expanded_g})
into Eq.~(\ref{cas_en}) we get an expression for the vacuum energy
\begin{equation}
  E=\frac{1}{2i}\int_{-\infty}^\infty\frac{d \omega}{2\pi} 2 \omega^2
  \int_{-\infty}^\infty\frac{dk_z}{2\pi}\int d z
  \sum_\eta \int_a^b \rho \,d \rho\, R_\eta^2(\rho;\omega,k_z)
  \int d\theta\left[g_\eta(\theta,\theta)-g_\eta^{(0)}(\theta,\theta)\right].
\end{equation}
We can simplify the result considerably
if we now make a Euclidean rotation from
$\omega\to i \zeta$,  make the substitution
$\zeta^2+k_z^2=\kappa^2$, and integrate out the angle in the
$\zeta$-$k_z$ plane to get the expression for the energy per unit length

\begin{equation}\label{e-rg}
  \mathcal{E}=-\frac{1}{4\pi}\int_0^\infty \kappa^3 d \kappa
  \sum_\eta \int_a^b \rho \,d \rho\, R_\eta^2(\rho;\kappa)
  \int d\theta\left[g_\eta(\theta,\theta)-g_\eta^{(0)}(\theta,\theta)\right].
\end{equation}

\subsection{The Radial Eigenvalue Problem}
We see that we need an expression for the radial integral
\begin{equation}\label{rad_int_unsolved}
  \int_a^b \rho \,d \rho\, R_\eta^2(\rho;\kappa),
\end{equation}
where the $R_\eta$s are the normalized eigenfunctions obeying the differential
equation (\ref{rad_de}). The normalization is
\begin{equation}
  \int_a^b \frac{d \rho}{\rho} R_\eta^2(\rho;\kappa)=1.
\end{equation}

To evaluate this integral we will use the identity (\ref{ident1}).
%\begin{equation}\label{cool_ident}
%  \frac\partial{\partial x}\left(p(x)\frac\partial{\partial x}u_\lambda(x)
%\frac\partial{\partial\lambda}v_\lambda(x)
%  -p(x)u_\lambda(x)\frac\partial{\partial\lambda}\frac\partial{\partial x}
%v_\lambda(x)\right)=  2\lambda r(x) u_\lambda(x) v_\lambda(x)
%\end{equation}
%where $u_\lambda(x)$ and $v_\lambda(x)$ are any two solutions to the
%differential equation
%\begin{equation}
%\left[-\frac{d}{dx}p(x)\frac{d}{dx} + q(x) -\lambda^2 r(x) \right]
%\psi_\lambda(x)=0.
%\end{equation}
The boundary conditions are that
$R_\eta(a;\kappa)=R_\eta(b;\kappa)=0$; this is only possible for
discrete values of $\eta$, namely, this is an eigenvalue condition
for $\eta$.
Let
$\tilde{R}_\eta(r;\kappa)$ be a solution to Eq.~(\ref{rad_de}) which satisfies
$\tilde{R}_\eta(a;\kappa)=0$ for all $\eta$ and $\kappa$. The normalized
solution can then be written as
\begin{equation}
  R_\eta(\rho;\kappa)=\frac{1}{N}\tilde{R}_\eta(\rho;\kappa),
\end{equation}
where
\begin{equation}
  N^2=\int_a^b\frac{d\rho} {\rho}\tilde{R}_\eta^2(\rho,\kappa).\label{norm}
\end{equation}
Now writing Eq.~(\ref{rad_de}) as
\begin{equation}
  \left[-\frac{d}{d\rho}\rho\frac{d}{d\rho}+\kappa^2 \rho -
    \eta^2 \left(\frac{1}{\rho}\right) \right]R_\eta(\rho,\kappa)=0,
\label{cdhere}
\end{equation}
we can see that $\frac{1}{\rho}\tilde{R}_\eta^2(\rho;\kappa)$ is a total
derivative given by Eq.~(\ref{ident1}).
(We replace $\kappa\to\eta$ there.) The integral (\ref{norm})
is now trivial to
carry out. We  see that the value at the lower limit of
integration is zero by our boundary condition that
$\tilde{R}_\eta(a;\kappa)=0$, and the second term on the right
in Eq.~(\ref{ident1}) at the upper limit is zero
by the eigenvalue condition
$\tilde{R}_\eta(b;\kappa)=0$. This gives the normalization constant as
\begin{equation}
  N^2=\frac{b}{2\eta}\frac\partial{\partial b}\tilde{R}_\eta(b;\kappa)
  \frac\partial{\partial\eta}\tilde{R}_\eta(b;\kappa).
\end{equation}

Now by considering $\kappa$ rather than $\eta$ as the parameter
in Eq.~(\ref{cdhere})
we also have from Eq.~(\ref{ident1}) that the desired integral
(\ref{rad_int_unsolved}) is a total derivative,
\begin{equation}
  \int_a^b\rho\, d \rho \,\tilde{R}_\eta^2(\rho,\kappa)= -\frac{b}{2\kappa}
  \frac\partial{\partial b}\tilde{R}_\eta(b;\kappa)
\frac\partial{\partial\kappa}\tilde{R}_\eta(b;\kappa).
\end{equation}
So the desired integral  given by Eq.~(\ref{rad_int_unsolved}) can be
concisely written as
\begin{equation}
  \int_a^b \rho\, d \rho\, R_\eta^2(\rho;\kappa)=-\frac{\eta}{\kappa}
  \frac{\frac\partial{\partial\kappa}\tilde{R}_\eta(b;\kappa)}
  {\frac\partial{\partial\eta}\tilde{R}_\eta(b;\kappa)}.
\end{equation}

\subsection{Argument Principle}
Now we again use the argument principle (\ref{arg-prin}),
which %gives a sum over non-explicit
%eigenvalues and turns it into a contour integral around the real line,
%\begin{equation}
%  \sum_\lambda f(\lambda)=\frac{1}{2\pi i}\int\limits_\gamma d\lambda
%  \left(\frac\partial{\partial\lambda}\ln D(\lambda)\right) f(\lambda).
%\label{arg-prin}
%\end{equation}
%The contour of integration $\gamma$ is illustrated in Fig.~\ref{fig-gamma}.
%\begin{figure}[tb]
%  \begin{center}
%  \includegraphics[width=3in]{fig8.eps}
%  \caption{Contour of integration $\gamma$ for the argument principle
%(\ref{arg-prin}).}
%  \label{fig-gamma}
%  \end{center}
%\end{figure}
we previously used for the angular eigenvalues;
in this case the sum is over the radial eigenvalues, and the
eigenvalue condition is given by $D(\eta)=R_\eta(b;\kappa)$. So we have
occuring in the energy (\ref{e-rg}) the form
\bea
  \sum_\eta \int_a^b \rho \,d \rho R^2_\eta(\rho;\kappa)&=&
\frac1{2\pi i}  \int\limits_\gamma d\eta
\frac{\frac\partial{\partial\eta}\tilde{R}_\eta(b;\kappa)}
  {\tilde{R}_\eta(b;\kappa)}\left(-\frac{\eta}{\kappa}
  \frac{\frac\partial{\partial\kappa}\tilde{R}_\eta(b;\kappa)}
  {\frac\partial{\partial\eta}\tilde{R}_\eta(b;\kappa)}\right)\nonumber\\
&=&  -\frac1{2\pi i}
\int_\gamma d\eta\frac{\eta}{\kappa}\frac\partial{\partial\kappa}
\ln \tilde{R}_\eta(b;\kappa).\eea
The
expression for the Casimir energy per length (\ref{e-rg}) is then given by
\begin{equation}
  \mathcal{E}=\frac{1}{8 \pi^2 i}\int_0^\infty \kappa^2 \,d \kappa
  \int\limits_\gamma \eta \,d \eta
  \left(\frac\partial{\partial\kappa}\ln \tilde{R}_\eta(b;\kappa) \right)
  \int d\theta\left(g_\eta(\theta,\theta)-g_\eta^{(0)}(\theta,\theta) \right).
\end{equation}

\subsection{The Radial Solutions}
The differential equation (\ref{cdhere})
%\begin{equation}
%  \left[-\rho\frac{d}{d\rho}\rho\frac{d}{d\rho}+\kappa^2\rho^2
%-\eta^2\right]R_\eta(\rho;\kappa)=0,
%\end{equation}
is the modified Bessel differential equation, of  imaginary
order. We need two independent solutions of this equation, which we
could take to be $K_{i \eta}(\kappa \rho)$ and $L_{i \eta}(\kappa \rho)$,
given by Eq.~(\ref{def-k-l}).
Now we want to find the solution $\tilde{R}_\eta(\rho;\kappa)$ that is
zero for $\rho=a$ for all values of $\eta$ and $\kappa$. An obvious solution is
\begin{equation}
  \tilde{R}_\eta(\rho;\kappa)=K_{i\eta}(\kappa a)\tilde I_{i\eta}(\kappa \rho) 
- \tilde I_{i\eta}(\kappa a) K_{i\eta}(\kappa \rho)=R_{-\eta}(\rho,\kappa),
\label{tilder}
\end{equation}
where
\be
\tilde I_\nu=\frac12(I_\nu+I_{-\nu})=\frac{\sin\nu\pi}{i\pi}L_\nu
\ee
is the function initially called $L_\nu$ in Ref.~\cite{ellingsen09};
here this is a more convenient choice, in that both $K_\nu$ and $\tilde I_\nu$
are even in $\nu$.

\subsection{Reduced Green's Function}

We also need the reduced Green's function in the angular
coordinates. The free Green's function is easily found to be
\begin{equation}
  g^{(0)}_\eta(\theta,\theta')=\frac{1}{2\eta}\left(-\sinh \eta|\theta-\theta'|
  +\frac{\cosh \eta \pi}{\sinh \eta \pi} \cosh \eta |\theta-\theta'|\right).
\end{equation}
If we assume a single $\delta$-function potential $v(\theta)=\lambda
\delta(\theta-\alpha)$ then the Green's function is
\begin{multline}\label{g_nu}
  g_\eta(\theta,\theta')=\frac{1}{2\eta}\bigg(-\sinh \eta |\theta-\theta'|
  +\frac{1}{2\eta \sinh \eta \pi +\lambda \cosh \eta \pi}\bigg[
    2\eta \cosh \eta \pi \cosh \eta |\theta-\theta'| \\
    - \frac{\lambda}{2\sinh \eta \pi} \Big\{
    \cosh \eta ( 2\pi +2 \alpha -\theta -\theta')
    -\cosh 2 \eta \pi \cosh \eta |\theta-\theta'|
    \Big\}\bigg] \bigg),
\end{multline}
which is defined for $\theta$ and $\theta'$ in the interval
$[\alpha,2\pi+\alpha]$.

The quantity of interest, $\tr(g-g^{(0)})$, is then
\begin{equation}
  \int_\alpha^{2\pi+\alpha}d\theta
\left[g_\eta(\theta,\theta)-g^{(0)}_\eta(\theta,\theta)\right]=
  \frac{-\lambda(\sinh \eta \pi \cosh \eta \pi + \eta \pi)}
  {2 \eta^2 \sinh \eta \pi ( 2\eta \sinh \eta \pi + \lambda \cosh \eta \pi)},
\end{equation}
and this expression can be seen to be a total derivative
\begin{equation}
  \int_\alpha^{2\pi+\alpha}d\theta
\left[g_\eta(\theta,\theta)-g^{(0)}_\eta(\theta,\theta)\right]=
\frac{1}{2\eta}\frac\partial{\partial\eta}
\ln \left(1+\frac{\lambda}{2\eta}\coth \eta \pi\right)=
\frac{1}{2\eta}\frac\partial{\partial\eta}
\ln \left(1+\lambda g_\eta^{(0)}(\alpha,\alpha)\right),
\end{equation}
which agrees with the result stated in Eq.~(\ref{d-1pot-e}).
It is precisely of the expected form (\ref{traceofg}).

\subsection{Casimir Energy}\label{sec4e}
The final form for the Casimir energy for a single radial
$\delta$-function potential in the annular region is
\bea
  \mathcal{E}&=&\frac{1}{16 \pi^2 i}\int_0^\infty\kappa^2d\kappa
  \int\limits_\gamma d \eta
  \left( \frac\partial{\partial\kappa} \ln \left[
    K_{i\eta}(\kappa a)\tilde I_{i\eta}(\kappa b) -
    \tilde I_{i\eta}(\kappa a) K_{i\eta}(\kappa b)\right]\right)\nonumber\\
 &&\qquad\times \left(\frac\partial{\partial\eta}\ln \left[1+\frac{\lambda}
    {2\eta}\coth \eta \pi \right] \right).\label{ce-1p}
\eea

This result may also be obtained by the multiple scattering
formalism \cite{Milton:2007wz}, which says that
\be
E=-\frac1{2i}\Tr \ln GG^{(0)}{}^{-1}=\frac1{2i}\Tr\ln(1+G^{(0)}V),
\ee
the latter form being a useful form for a single potential.  We see
that Eq.~(\ref{ce-1p}) exactly corresponds to this if we integrate
by parts on $\eta$ and $\kappa$:
\be
\mathcal{E}=\frac1{8\pi^2i}\int_0^\infty d\kappa\,\kappa\int_\gamma d\eta
\left(\frac{\partial}{\partial\eta}\ln \left[
    K_{i\eta}(\kappa a)\tilde I_{i\eta}(\kappa b) -
    \tilde I_{i\eta}(\kappa a) K_{i\eta}(\kappa b)\right]\right)
\ln\left[1+\lambda g^{(0)}(\alpha,\alpha) \right].\label{ce-1p-ms}
\ee

\begin{figure}[tb]
  \begin{center}
    \includegraphics[width=5in]{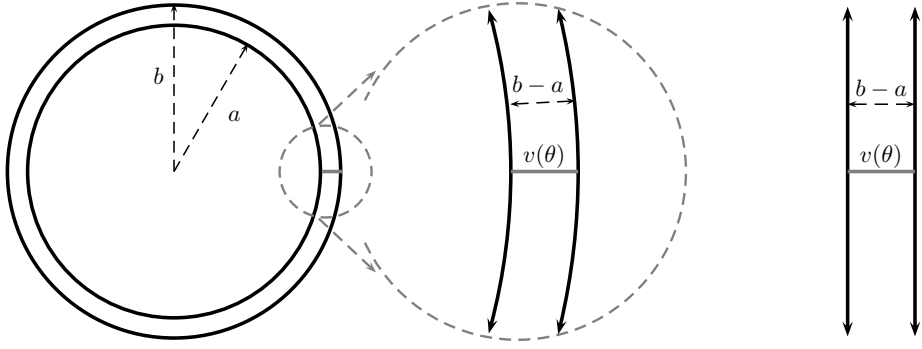}
    \caption{\label{fig-lim} Large radius limit of annular geometry.
For large annular radii $a$ and $b$, with $b-a$ fixed, the annular
boundaries become indistinguishable from parallel planes.}
  \end{center}
\end{figure}

We can check this result by taking the limit as $a$
and $b$ get very large, but with fixed distance between the circles.
In this limit this result should reproduce the
case of a single semitransparent plane between two parallel Dirichlet
planes, as illustrated in Fig.~\ref{fig-lim}.
The energy in that case should be
\begin{equation}
  \mathcal{E}=-\frac{1}{8 \pi} \int_0^\infty \kappa^3 d \kappa
  \sum_{n=1}^\infty
  \frac{1}{\tilde\eta}\frac\partial{\partial\tilde\eta}
  \ln\left(1+\frac{\tilde\lambda}{2\tilde\eta}\right),
\end{equation}
where $\tilde\eta^2 =\kappa^2+(n \pi / (b-a) )^2$. If we use the
argument principle we can write it as
\begin{equation}\label{ce-para}
  \mathcal{E}=-\frac{1}{16 \pi^2 i} \int_0^\infty \kappa^3 d \kappa
  \int\limits_\gamma d \tilde\eta
  \left( \frac\partial{\partial\tilde\eta} \ln \left[
\frac{\sin\sqrt{\tilde\eta^2-\kappa^2}
(b-a)}{\sqrt{\tilde\eta^2-\kappa^2}} \right]\right)
  \left( \frac1{\tilde\eta}\frac\partial{\partial\tilde\eta} \ln \left[
    1+\frac{\tilde\lambda}{2\tilde\eta} \right] \right).
\end{equation}
The square root divided out in the logarithm is present to remove
the spurious square-root singularity.
It should be noted that both of these expressions (\ref{ce-1p}) and
(\ref{ce-para}) are divergent, but
the divergence is simply the self-energy divergence always present with a
single plane.

It is straightforward to prove that Eq.~(\ref{ce-1p}) reduces to
Eq.~(\ref{ce-para}) in the appropriate limit.  The second logarithm
in the former becomes, in the limit $\eta\to\infty$, simply
$\ln(1+\lambda/2\eta),$ so that suggests the correspondence
\be
\tilde\eta=\frac{\eta}{a},\quad \tilde\lambda=\frac\lambda{a}.
\label{subs}
\ee
And the leading uniform asymptotic expansion of the modified
Bessel functions \cite{dunster90,olver} gives
\be
K_{i\eta}\left(\eta \frac{\kappa a}\eta\right)
\tilde I_{i\eta}\left(\eta \frac{\kappa b}\eta\right)-
K_{i\eta}\left(\eta \frac{\kappa b}\eta\right)
\tilde I_{i\eta}\left(\eta \frac{\kappa a}\eta\right)\sim
\frac1{2\eta}t(z_a)^{1/2}t(z_b)^{1/2}\sin
\left[\eta(f(z_a)-f(z_b))\right],
\ee
where
$z_a=\kappa a/\eta$, $z_b=\kappa b/\eta$, $t(z)=(1-z^2)^{-1/2}$, and
\be
f(z)=\ln\left(\frac{1+t(z)^{-1}}{z}\right)-\frac1{t(z)},
\quad f'(z)=-\frac1{z t(z)}.\ee
(The latter function is a continuation of a function
 usually called $\eta$, but we have already
used that symbol repeatedly.) This result holds true for $z<1$,
but an equivalent form, obtained by analytic continuation, holds
for $z>1$. (See Appendix \ref{app:b}.)
Then the derivative with respect to $\kappa$ term in
Eq.~(\ref{ce-1p}) becomes
\be
\frac\partial{\partial\kappa}\ln(K\tilde I-\tilde I K)
\sim\frac\partial{\partial\kappa}
\ln\left[\frac{1}{\sqrt{\eta^2-(\kappa a)^2}}\sin\left(
\sqrt{\eta^2-(\kappa a)^2}\frac{b-a}b\right)\right],
\ee
so with the substitutions (\ref{subs}), and the observation that
\be
\frac\partial{\partial\eta}F(\eta^2-\kappa^2)=-\frac\eta\kappa
\frac\partial{\partial\kappa}F(\eta^2-\kappa^2),
\ee
 we exactly recover the Casimir energy
for a semitransparent plate between two Dirichlet plates (\ref{ce-para}).

\begin{figure}[tb]
  \begin{center}
  \includegraphics[width=3in]{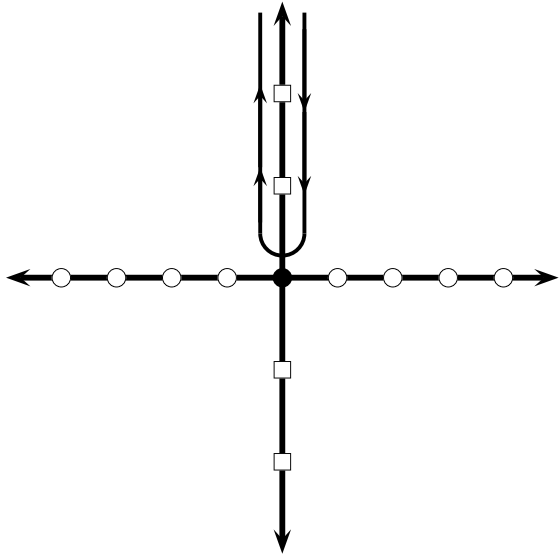}
  \caption{Transformed contour of integration for the $\nu$ integral in
Eq.~(\ref{e-conv-1p}).}
  \label{fig-Gamma}
  \end{center}
\end{figure}

It is also easy to check that the energy (\ref{ce-1p}) agrees
with the expression for the energy given by the more conventional
approach described in Sec.~\ref{sec-st}.  The latter is
\bea
\mathcal{E}&=&-\frac1{16\pi^2 i}\int_0^\infty \kappa^2\,d\kappa\int_\gamma
d\nu\left(\frac{\partial}{\partial\nu}\ln\left[1-
\frac\lambda{2\nu}\cot\nu\pi\right]\right)
\nonumber\\
&&\qquad\times\frac\partial{\partial \kappa}\ln\left[I_\nu(\kappa a)
K_\nu(\kappa b)-I_\nu(\kappa b)K_\nu(\kappa a)\right].\label{e-conv-1p}
\eea
This equivalence
may be easily shown by seeing that the integrand is odd in $\nu$,
because
\be I_\nu(\kappa a)K_\nu(\kappa b)-I_\nu(\kappa b)K_\nu(\kappa a)
=\tilde I_\nu(\kappa a)
K_\nu(\kappa b)-\tilde I_\nu(\kappa b)K_\nu(\kappa a),
\ee
and then rotating the contour $\gamma$
from that shown in Fig.~\ref{fig-gamma} to
that in Fig.~\ref{fig-unfold}, which may then be transformed to that shown
in Fig.~\ref{fig-Gamma}, by changing $\nu$ to $-\nu$ on the negative imaginary
axis.  Thus the contour $\gamma$ in the $\nu$ plane is transformed to $\gamma$
in the $\eta$ plane appearing in Eq.~(\ref{ce-1p}) (except traversed
in the opposite sense), and
the equivalence is established.

\subsection{Interaction Between Two Semitransparent Planes}
\begin{figure}[tb]
  \begin{center}
    \includegraphics[width=1.7in]{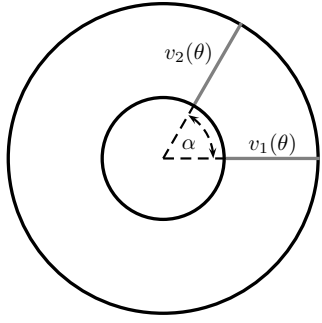}
    \caption{\label{fig-2-ann} Two semitransparent plates in an annulus.}
  \end{center}
\end{figure}

If we want to look at an explicitly finite quantity we will need to
look at the interaction energy between two semitransparent planes.
The geometry is illustrated in Fig.~\ref{fig-2-ann}.
We will use a slightly different form of the energy for this, based on the
multiple-scattering formalism \cite{Milton:2007wz}:
\begin{equation}
  E=\frac{1}{2i}\int_{-\infty}^\infty \frac{d \omega}{2\pi}
  \Tr \ln ( 1- \mathcal{G}^{(1)}V_1\mathcal{G}^{(2)}V_2 ).
\end{equation}
The subscripts on the $V$s represent the potentials $V_1(\mathbf{r})=
\lambda_1\delta( \theta ) / \rho^2$, and $V_2(\mathbf{r}) =
\lambda_2 \delta( \theta- \alpha ) / \rho^2$.
The Green's functions with superscript $(i)$
represent the interaction with only a single
potential $V_i$. By using Eq.~(\ref{expanded_g}), we can greatly simplify the
interaction energy to
\begin{equation}
  \mathcal{E}=\frac{1}{4\pi}\int_0^\infty \kappa \,d \kappa
  \sum_\eta \ln \left( 1- \tr g_\eta^{(1)}v_1 g_\eta^{(2)}v_2 \right).
\label{gvgv}
\end{equation}
We already have an expression for $g_\eta^{(i)}$, given in
Eq.~(\ref{g_nu}). Using the latter we can write
\begin{equation}
  \tr g_\eta^{(1)}v_1 g_\eta^{(2)}v_2 = \frac{\lambda_1 \lambda_2
    \cosh^2 \eta (\pi-\alpha)}
  {\left(2\eta \sinh \eta \pi +\lambda_1 \cosh \eta \pi\right)
    \left(2 \eta \sinh \eta \pi +\lambda_2 \cosh \eta \pi \right) }.
\end{equation}  This exactly agrees with Eq.~(\ref{fulldee-e}).

Using the argument principle to replace the sum we then get
the Casimir energy of two semitransparent plates in a Dirichlet annulus,
the immediate generalization of the Casimir energy (\ref{ce-1p-ms})
for a single plate,
\begin{multline}\label{e-2plates}
  \mathcal{E}=\frac{1}{8\pi^2 i}\int_0^\infty \kappa \,d \kappa
  \int\limits_\gamma d \eta
  \left( \frac\partial{\partial\eta} \ln \left[
    K_{i\eta}(\kappa a)\tilde I_{i\eta}(\kappa b) -
    \tilde I_{i\eta}(\kappa a) K_{i\eta}(\kappa b)\right]\right)\\
  \times\ln \left( 1- \frac{\lambda_1 \lambda_2 \cosh^2 \eta (\pi-\alpha)}
  {\left(2\eta \sinh \eta \pi +\lambda_1 \cosh \eta \pi\right)
    \left(2 \eta \sinh \eta \pi +\lambda_2 \cosh \eta \pi \right) } \right).
\end{multline}

A limiting case when $a$, $b\to\infty$, $b-a$ fixed,
should be two perpendicular semitransparent planes, a
distance $d$ apart, sandwiched between Dirichlet planes,
similar to the single plate situation treated in
the subsection above. A similar formula
should then be
\begin{equation}
  \mathcal{E}=\frac{1}{8\pi^2 i}\int_0^\infty \kappa \,d \kappa
  \int\limits_\gamma d \tilde\eta
  \left( \frac\partial{\partial\tilde\eta} \ln \left[ \frac{\sin
\left(\sqrt{\tilde\eta^2-\kappa^2} (b-a)\right)}{\sqrt{\tilde\eta^2-\kappa^2}}
\right] \right)
  \ln \left(1 -\frac{\tilde\lambda_1\tilde\lambda_2 e^{-2 \tilde\eta d }}
  {(2\tilde\eta+\tilde\lambda_1)(2\tilde\eta+\tilde\lambda_2)}\right).
\end{equation}
As in the case of a single plate, this limiting form is immediately obtained
from Eq.~(\ref{e-2plates}).

Finally, we verify that we obtain the expression (\ref{expression})
for the wedge geometry.
To do this, we must include the modes exterior to the outer cylinder
(with the wedge extended to infinity as in Fig.~\ref{fig1})
and subtract the energy present if the outer cylinder were not present.
This means that the radial dispersion function determining the
azimuthal eigenvalues $\eta$ becomes
\be
\hat R_\eta(b;\kappa)=
\left(K_{i\eta}(\kappa a)\tilde I_{i\eta}(\kappa b)
-\tilde I_{i\eta}(\kappa a)K_{i\eta}(\kappa b)
\right)
\frac{K_{i\eta}(\kappa b)}{K_{i\eta}(\kappa a)}.
\ee
The extended annular energy is then given by Eq.~(\ref{e-2plates}) with
$\tilde R_{\eta}(b;\kappa)\to \hat R_{\eta}(b;\kappa)$.
We now can distort the contour $\gamma$ to one lying along the imaginary
axis as shown in Fig.~\ref{fig-unfold}, $i\eta\to\nu$,
%\begin{figure}[tb]
%  \begin{center}
%    \includegraphics[width=1.7in]{fig11.eps}
%    \caption{\label{fig11}Distortion of the contour $\gamma$ into
%one lying along the imaginary $\nu$ axis.}
%  \end{center}
%\end{figure}
(because the second logarithm in Eq.~(\ref{e-2plates}) falls off
exponentially fast for $\re\eta>0$), and then using the small argument
expansion, for real $\nu$,
\be
K_\nu(x)\sim \frac{\Gamma(|\nu|)}2\left(\frac{x}2\right)^{-|\nu|},
\quad \tilde I_\nu(x)\sim \frac{\sin|\nu|\pi}{\pi}\frac{\Gamma(|\nu|)}2
\left(\frac{x}2\right)^{-|\nu|}, \quad x\to 0.
\ee
This means for small $a$ and real $\nu$
 the first logarithm in Eq.~(\ref{e-2plates}) is
$\ln I_\nu(\kappa b)K_\nu(\kappa b)$,
which is just what was encountered in Eq.~(\ref{intofg}).
We then fold the $\nu$ integral to encircle
 the positive real axis as in Fig.~\ref{fig-Gamma} and integrate
by parts in $\kappa$ and $\nu$.
In this way  the form (\ref{eform}) is reproduced (with $D\to\hat D$), 
which
leads to the final expression (\ref{expression}).

\subsection{Numerical Evaluation of the Casimir Energy for Two 
Dirichlet Planes in an Annulus}
\begin{figure}[hb]
\includegraphics{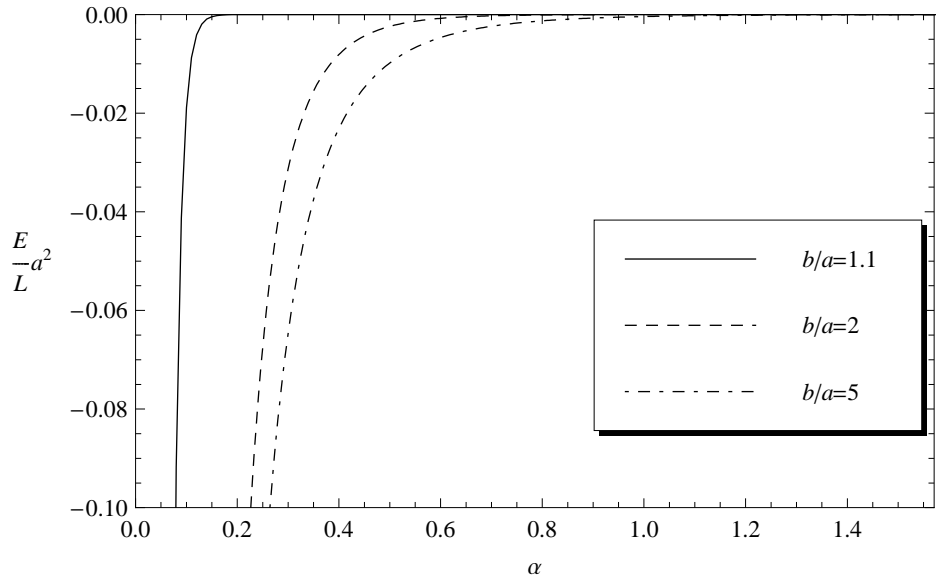}
\caption{\label{ann:pi_4}This figure shows the energy per length vs
  the angle between Dirichlet plates. The energy is scaled by the inner
  radius $a$. The three lines represent three different ratios of inner
  to outer radius $\frac{b}{a}=1.1$, $\frac{b}{a}=2$, and
  $\frac{b}{a}=5$.}
\end{figure}

The Casimir energy in  Eq.~\eqref{gvgv} is a quickly converging
function so it should be easy to evaluate. However, it can be difficult
to evaluate the $\eta$ eigenvalues, which become functions of the
wavenumber $\kappa$ and a natural number $m$. We can get around this
problem, again, by exploiting the argument principle in order to get a
contour integral in the complex plane, as in Eq.~\eqref{e-2plates}. We cannot
integrate along the real line because of the poles introduced when we
use the argument principle, and unlike with the wedge we cannot open
along the imaginary axis, because the integral then becomes
divergent. So a simple choice is then to let the integration run along
the angles of $\pi/4$ and $-\pi/4$
in the complex $\eta$ plane. Identifying $R_\eta(b,\kappa)$
from Eq.~\eqref{tilder}, and writing 
$\tr g_\eta^{(1)}v_1g_\eta^{(2)}v_2=A(\eta)$
we have
\begin{multline}
\mathcal{E}=\frac{1}{4\pi^2}
\int_0^\infty \kappa d \kappa \int_0^\infty \dif\nu\Bigg\{
\frac{\Re R_{\sqrt{i}\nu}\partial_\nu\Re R_{\sqrt{i}\nu}+
\Im R_{\sqrt{i}\nu}\partial_\nu\Im R_{\sqrt{i}\nu}}{
\left|R_{\sqrt{i}\nu}\right|^2}
\arctan\left(\frac{\Im A(\sqrt{i}\nu)}{1-\Re A(\sqrt{i}\nu)}\right)\\
-\frac{\Re R_{\sqrt{i}\nu}\partial_\nu\Im R_{\sqrt{i}\nu}-
\Im R_{\sqrt{i}\nu}\partial_\nu\Re R_{\sqrt{i}\nu}}{2
\left|R_{\sqrt{i}\nu}\right|^2}
\ln\left(1-2 \Re A(\sqrt{i}\nu)+\left|A(\sqrt{i}\nu)\right|^2\right)\Bigg\}.
\end{multline}
Here we have used the property that $R_{\eta^*}=R_\eta^*$, and
$A(\eta^*)=A^*(\eta)$. The value of $R_{\sqrt{i}\nu}(b,\kappa)$ is
obtained as a numerical solution to the differential
equation. Using this technique we can obtain a numerical energy in
about 1 cpu-second. The results of this calculation are found in
Fig.~\ref{ann:pi_4}.

\begin{figure}[hb]
\includegraphics{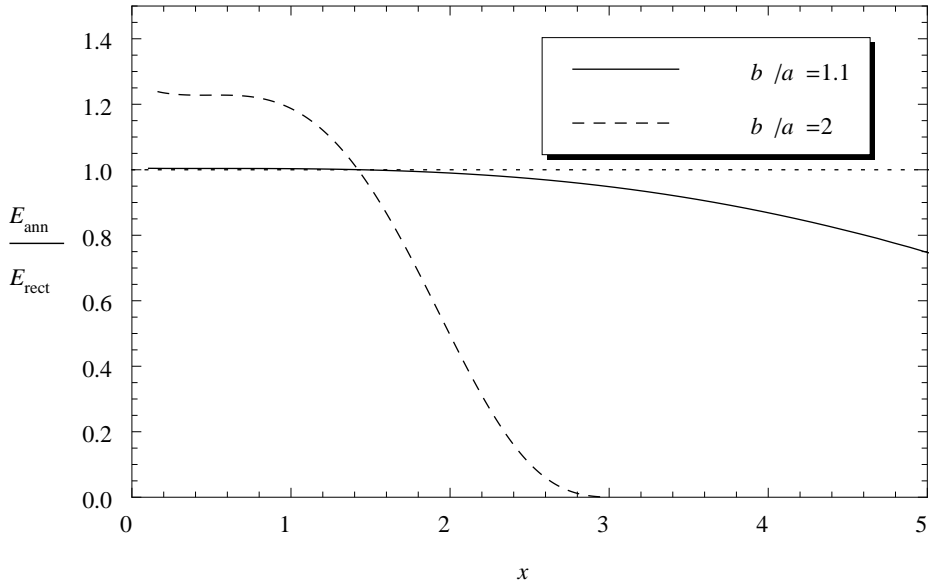}
\caption{\label{ann:comparison}This figure shows the ratio of the
  energies of an annular Dirichlet piston to a rectangular 
Dirichlet piston of similar
  dimension vs average separation distance between the plates. The variable
$x$ is the  separation distance  scaled by the finite size of the piston
  $b-a$, $x=d/(b-a)$. 
The two lines represent two ratios of inner to outer radius
  $\frac{b}{a}=1.1$, and $\frac{b}{a}=2$.  For the latter case,
only the region $\alpha\in[0,\pi]$ is shown.}
\end{figure}

Again we would like to compare to known results, so
Fig.~\ref{ann:comparison} is a graph of the ratio of the energies of an
annular piston to a rectangular piston of similar dimension. The
rectangular piston is constructed so it has the same finite width
$b-a$ as the annular piston, and the separation distance is the mean
distance between the annular plates,
\begin{equation}
d=\frac{b+a}{2}2\sin\left(\frac{\alpha}{2}\right).
\end{equation}
The results make a certain amount of physical sense. The energy of the
annular piston is greater than that of the rectangular piston for
small separation because the inner edge of the annular piston is
closer, and will contribute more to the energy. However as the annular
piston gets further away, the other side of the piston will start to
contribute and lower the overall energy. In addition we see that the
small ratio piston is much closer to the rectangular piston for small
separations than a larger ratio, and in
the plateau region for small separations, 
$E_{\text{ann}}/E_{\text{rect}}\approx
  1.04$ for $b/a=1.1$ vs. $E_{\text{ann}}/E_{\text{rect}}\approx
  1.23$ for $b/a=2$. These numbers are quite closely reproduced by
the ratio of the proximity force approximate value of the energy
for tilted plates to the energy for parallel plates (ignoring the
sidewalls) for small tilt angles,
\be \frac{\mathcal{E}_{\rm PFA}}{\mathcal{E}_{\|}}=\frac1{16}\frac{a^2}{b^2}
\left(1+\frac{a}b\right)^4.
\ee

\section{Theta dependent potentials}
\label{sec:theta}
Instead of considering, as usually done,
spherically or cylindrically symmetric
potentials, in this paper we have been examining potentials depending on the
angles. In particular, in two dimensions, in order for separation of variables
to work, we have been considering
the operator
\be
L=-\nabla^2 + \frac 1 {\rho^2} v(\theta ) = - \frac{\partial^2}
{\partial \rho^2} - \frac 1 \rho
\frac \partial {\partial \rho} - \frac 1 {\rho^2} \frac{ \partial^2}
{\partial \theta ^2} + \frac 1 {\rho^2} v(\theta ) , \ee
with $v(\theta )$ given in Eq.~(\ref{dfpot}). The advantage 
of this potential is 
that a closed form for the secular equation can be given, see Eq.~(\ref{a6}). 
But what can be said for other potentials $v(\theta )$? 
The relevant equations for the two-dimensional 
Green's function are still Eqs.~(\ref{twogr})-(\ref{gradcir}). 
In particular, for the angular eigenfunctions we still have
%Eigenvalues $\lambda^2$ and eigenfunctions $u_\lambda (\rho,\theta )$
%are defined by
%\be L u_\lambda (\rho, \theta ) = \lambda^2 u_\lambda (\rho,\theta),
%\label{original}
%\ee
%with suitable boundary conditions imposed. Assuming eigenfunctions of the form
%$u_\lambda (\rho,\theta ) = R(\rho) \Theta (\theta )$, the separated equations
%read
\bea 0 &=& \Theta'' (\theta ) + (\nu^2 - v (\theta ) ) \Theta (\theta ),
\label{septheta}\eea
%0 &=& R'' (\rho) + \frac 1 \rho R' (\rho) +
%\left(\lambda^2 - \frac{\nu^2} {\rho^2} \right)
%R(\rho) , \label{sepr} \eea
where $\nu$ is the separation constant. The separation constant is determined
from the boundary condition in $\theta$. 
For $v(\theta )$ a smooth potential one imposes periodic boundary
conditions, and this is what we concentrate on for concreteness.

Note, that the only information from Eq.~(\ref{septheta}) that enters
Eq.~(\ref{redgre}) for the radial reduced Green's function is 
the separation constant $\nu$. From Eq.~(\ref{septheta})
its square can be considered the eigenvalue of
\be
L_\theta = -
\frac{ \partial^2} {\partial \theta ^2} + v(\theta )\ee with periodic boundary
conditions. For a nontrivial potential $v(\theta )$ no explicit form of $\nu$
will be known. But also in general a transcendental equation determining the
eigenvalues can be obtained; we follow Ref.~\cite{kirs04-37-4649}. In order to
formulate this equation let $H(\theta )$ be the fundamental matrix of
Eq.~(\ref{septheta}). That is, let $u_\nu ^{(1)} (\theta )$ and $u_\nu ^{(2)}
(\theta )$ be two linearly independent solutions of Eq.~(\ref{septheta}). With
$w_\nu^{(i)} (\theta )= du_\nu^{(i)} (\theta ) / d\theta$, the fundamental
matrix is
\bea H (\theta )= \left( \begin{array}{cc} u_\nu ^{(1)} (\theta )
& u_\nu ^{(2)} (\theta ) \\
w_\nu ^{(1)} (\theta ) & w_\nu ^{(2)} (\theta ) \end{array} \right),\eea
where we choose the normalizations such that $H(0) = 1_{2\times 2}.$ With
these definitions and normalizations,
%let $u_\nu (\theta )$ be a solution to (\ref{septheta})
%with {\it initial value}
%\be {u_\nu (0) \choose w_\nu (0) } = { u_\nu ^{(2)} (2\pi)
%\choose 1-u_\nu ^{(1)} (2\pi) } .
%\ee
%In terms of the fundamental matrix we then have
%\be {u_\nu (\theta )
%\choose w_\nu (\theta ) } = H_\nu (\theta ) { u_\nu (0) \choose w_\nu (0)}
%\ee and
the equation determining the eigenvalues reads
\be  0=D(\nu )=(1-u_\nu^{(1)} (2\pi )) (1-w_\nu^{(2)} (2\pi )) - u_\nu ^{(2)} (2\pi ) w_\nu ^{(1)} (2\pi ). \label{perdnu}\ee
The solutions to this equation have to be used in Eq.~(\ref{redgre}).
%Solutions regular
% at the origin are $J_\nu (\lambda \rho)$ and
%eigenvalues are determined from the
%boundary condition in $\rho$.
%For example, with Dirichlet boundary conditions at
%$\rho=a$, eigenvalues follow from
%\be
%J_\nu (\lambda a ) =0.
%\ee
%By using the argument principle for the $\nu$ as well as the
%$\lambda$-summation, the zeta function for this setting therefore reads
%\be \zeta (s) = \frac 1 {2\pi i} \int\limits_\gamma d\nu
%\left\{\frac 1 {2\pi i}
% \int\limits_\Gamma dz \,\, z^{-2s} \frac d {dz} \ln J_\nu
%(za) \right\} \frac d {d\nu} \ln \left( w_\nu (0) - w_\nu (2\pi) \right) ,
%\label{zetadoucon}\ee
%for suitable contours $\gamma$ and $\Gamma$.
%The correspondence with Eqs.~(\ref{eform})--(\ref{intofg}) is now rather
%evident.
The Casimir energy expressions (\ref{eform}) and (\ref{intofg}) then 
remain valid, once the appropriate radial reduced Green's function is used
and $D(\nu )$ given in Eq.~(\ref{perdnu}) is substituted. 
Once $v(\theta )$ is specified this allows, in principle, for a numerical 
evaluation of the Casimir energy when suitable subtractions are performed.

%%%%%%%%%%%%%%%%%% ACKNOWLEDGEMENTS  %%%%%%%%%%%%%%%%%%%
%\subsection*{Acknowledgements}
\acknowledgments
This material is based upon work supported by the
National Science Foundation under Grants Nos.~PHY-0554926 (OU) and
PHY-0757791 (BU) and by the
US Department of Energy under Grants Nos.~DE-FG02-04ER41305 and DE-FG02-04ER-%
46140 (both OU).
We thank Simen Ellingsen, Iver Brevik, Prachi Parashar,
Nima Pourtolami, and Elom Abalo for collaboration.
Part of the work was done while KK enjoyed the hospitality and partial 
support of the
Department of Physics and Astronomy of the University of Oklahoma. 
Thanks go in particular to Kimball Milton and his group who
made this very pleasant and exciting visit possible.

\appendix
\section{An Integral Theorem}\label{sec4}
It may be useful to see explicitly how the trace of the subtracted
reduced Green's function turns into a derivative of a logarithm,
as in Eq.~(\ref{intofg}).  Consider a Green's function $g_\kappa(x,x')$
for a one-dimensional problem described by the differential equation
\be
\left[-\frac{d}{dx}p(x)\frac{d}{dx}-\kappa^2 r(x)+q(x)+p(x)V(x)\right]
g_\kappa(x,x')=\delta(x-x'),
\ee
where $V$ is a $\delta$-function potential,
\be
V(x)=\lambda\delta(x-c).
\ee
The problem is defined on the interval $a<c<b$, where at the boundaries
$g_\kappa$ satisfies Dirichlet boundary conditions,
\be
g_\kappa(a,x')=g_\kappa(b,x')=0.
\ee
If the potential $V=0$, let the corresponding Green's function be denoted
by $g_\kappa^{(0)}$.

Let us solve this problem in terms of two independent solutions of the
homogeneous equation
\be
\left[-\frac{d}{dx}p(x)\frac{d}{dx}-\kappa^2 r(x)+q(x)\right]
u_\kappa(x)=0.\label{dehomo}
\ee
Let $A_\kappa$ be such a solution that vanishes at the left boundary,
$A_\kappa(a)=0$,
and $B_\kappa$ be an independent solution that vanishes at the right boundary,
$B_\kappa(b)=0$, and let them be normalized so that the Wronskian is
\be
W[A_\kappa,B_\kappa](x)\equiv A_\kappa(x)B'_\kappa(x)-B_\kappa(x)A'_\kappa(x)
=-\frac1{p(x)}.
\ee
Then the ``free'' Green's function is
\be
g_\kappa^{(0)}(x,x')=A_\kappa(x_<)B_\kappa(x_>),
\ee
and the full Green's function has the form
\be
g_\kappa(x,x')=g_\kappa^{(0)}(x,x')+\left\{\begin{array}{cc}
\alpha A_\kappa(x)A_\kappa(x'),&a<x,x'<c,\\
\beta B_\kappa(x)B_\kappa(x'),&c<x,x'<b.\end{array}\right.
\ee
Now, it is easy to prove that
\bea
\alpha=\frac{\lambda B^2_\kappa(c)}{\lambda A_\kappa(c)B_\kappa(c)+1},\quad
\beta=\frac{\lambda A^2_\kappa(c)}{\lambda A_\kappa(c)B_\kappa(c)+1}.
\eea

It is immediate that any two solutions of the differential equation
(\ref{dehomo}) $u_\kappa$ and $w_\kappa$ satisfy
\be
\frac{\partial}{\partial x}\left[p(x)\left(
\frac{\partial}{\partial x}u_\kappa(x)\frac{\partial}{\partial\kappa}
w_\kappa(x)
-u_\kappa(x)\frac{\partial}{\partial\kappa}
\frac{\partial}{\partial x}w_\kappa(x)\right)\right]=2 \kappa
r(x)u_\kappa(x) w_\kappa(x),\label{ident1}
\ee
and therefore the following indefinite integral follows:
\be
\int dx\, r(x)u_\kappa^2(x)=\frac{p(x)}{2\kappa}u_\kappa(x)u'_\kappa(x)
\frac{\partial}{\partial\kappa}\ln\frac{u_\kappa(x)}{u'_\kappa(x)},
\ee
where $u'_\kappa(x)\equiv\frac{\partial}{\partial x}u_\kappa(x)$.

Now we can evaluate the trace of the interaction part of the Green's
function,
\bea
\mbox{tr}\,(g-g^{(0)})&\equiv& \int_a^b dx\,  r(x)\left[g(x,x)-g^{(0)}(x,x)
\right]\nonumber\\
&=&\frac{\lambda}{\lambda A_\kappa(c)B_\kappa(c)+1}\left[B^2_\kappa(c)
\int_a^c dx\,r(x)A_\kappa^2(x)+A_\kappa^2(c)\int_c^b dx\, r(x)
B_\kappa^2(x)\right]\nonumber\\
&=&-\frac{p(c)}{2\kappa}\frac{\lambda A_\kappa^2(c)B_\kappa^2(c)}{\lambda
A_\kappa(c)B_\kappa(c)+1}
\left[\frac{A_\kappa'(c)}{A_\kappa(c)}\frac{d}{d\kappa}
\ln\frac{A_\kappa'(c)}{A_\kappa(c)}-\frac{B_\kappa'(c)}{B_\kappa(c)}
\frac{d}{d\kappa}\ln\frac{B_\kappa'(c)}{B_\kappa(c)}\right]\nonumber\\
&=&\frac{d}{d\kappa^2}\ln\left[1+\lambda A_\kappa(c)B_\kappa(c)\right]
=\frac{d}{d\kappa^2}\ln\left[1+\lambda g^{(0)}(c,c)\right].
\label{traceofg}
\eea
This is the expected expression. As shown in Sec.~\ref{sec4e}, this is just
the expected multiple scattering result.
 In the Dirichlet limit $\lambda\to \infty$,
this agrees with the Bessel function result (\ref{intofg}), where
$p(x)=r(x)=x$; although in that case the boundary condition is not Dirichlet
at the origin, $p(0)=0$.

\section{Modified Bessel Functions of Pure Imaginary Order}\label{app:b}
In this work we encountered the following differential equation
\begin{equation}
  \left(x\frac\partial{\partial x}x\frac{\partial}{\partial x}-
x^2+\eta^2\right)\psi(x)=0,
\end{equation}
which is the modified Bessel equation, with the wrong sign for the
order parameter $\eta^2$. The solutions are then obviously modified
Bessel functions of imaginary order. So we might choose as the
independent pair of solutions the modified Bessel function of the first
kind, of positive and negative pure imaginary order $I_{i\eta}(x)$ and
$I_{-i\eta}(x)$. However the $I_{i\eta}$'s are not numerically
satisfactory functions; their values for real $x$ are complex, and the
phase is $x$ dependent. A standard pair of functions
can be defined as
\begin{subequations}
\bea
  K_\nu(x)&=&\frac{\pi}{2\sin \nu \pi}\left(I_{-\nu}(x)-I_\nu(x)\right),
\\
  L_\nu(x)&=&\frac{i \pi}{2\sin \nu \pi}\left(I_\nu(x)+I_{-\nu}(x)\right).
\label{lnu}
\eea
\end{subequations}
The $K_\nu(x)$ is the standard modified Bessel function of the second
kind, also called the Macdonald function. Both $K_{i\eta}(x)$ and
$L_{i\eta}(x)$ are real for real values of $\eta$ and $x$. For a fixed
$\eta$, both $K$ and $L$ oscillate with relatively constant amplitude
for $x<\eta$ and they die or grow exponentially for $x>\eta$
respectively. The limiting behaviors are given in this appendix;
see Refs.~\cite{dunster90,olver}.  Although this definition of
$L_\nu$ is convenient in Sec.~\ref{sec-st}, for the considerations
of Sec.~\ref{sec:alt}, the $\sin\nu\pi$ in Eq.~(\ref{lnu}) introduces
spurious singularities, and it is more convenient there to simply
use
\be
\tilde I_\nu(x)=\frac12\left(I_\nu(x)+I_{-\nu}(x)\right)=\frac{\sin\nu\pi}
{i\pi}L_\nu(x),
\ee
also called $L_\nu$ in Ref.~\cite{ellingsen09}.  In the following we will
give the behaviors of $K_{i\eta}$ and $\tilde I_{i\eta}$.

\subsection{Small Argument}
For fixed $\eta>0$, in the limit as $x\to0^+$,
\begin{subequations}
\bea
  K_{i \eta}(x)&=&-\left(\frac{\pi}{\eta \sinh \eta \pi}\right)^{1/2}
  \left[\sin\left( \eta \ln \frac{x}{2}-\phi_{\eta}\right)+
    \mathcal{O}(x^2)\right],
\\
  \tilde I_{i \eta}(x)&=&\left(\frac{\sinh\eta\pi}{\eta  \pi}\right)^{1/2}
  \left[\cos\left( \eta \ln \frac{x}{2}-\phi_{\eta}\right)+
    \mathcal{O}(x^2)\right],
\eea\end{subequations}
where $\phi_\eta$ is given by
\begin{equation}
  \phi_\eta=\arg [\Gamma(1+i \eta)].
\end{equation}

\subsection{Large Argument}
For fixed $\eta>0$ and large argument $|x|\to\infty$ we have
\begin{subequations}
\bea
  K_{i \eta}(x)&=&\left(\frac{\pi}{2 x}\right)^{1/2}e^{-x}
  \left[1+\mathcal{O}(x^{-1})\right],
  \qquad |\arg x| \le \frac{3\pi}{2}-\delta,
\\
  \tilde I_{i\eta}(x)&=&
  \left(\frac1{2\pi x}\right)^{1/2}e^{x}
  \left[1+\mathcal{O}(x^{-1})\right],
  \qquad |\arg x| \le \frac{\pi}{2}-\delta,
\eea
\end{subequations}
for arbitrary $\delta>0$.

\subsection{Uniform Asymptotic Expansion}
The uniform asymptotic expansions are for both large order and
argument. For fixed $z>0$ we have for the leading behavior
\begin{subequations}
\bea
  K_{i \eta}(\eta z)&\sim& \frac{\pi e^{-\eta \pi /2}}{\eta^{1/3}}
  \left(\frac{4 \zeta}{1-z^2}\right)^{1/4} \Ai(-\eta^{2/3}\zeta),
\\
  \tilde I_{i\eta}(\eta z)&\sim& \frac{e^{\eta \pi /2}}{2\pi\eta^{1/3}}
  \left(\frac{4 \zeta}{1-z^2}\right)^{1/4} \Bi(-\eta^{2/3}\zeta),
\eea
\end{subequations}
where the $\Ai(x)$ and $\Bi(x)$ are the Airy functions of the first
and second kinds respectively, and $\zeta$ is given by the relation
\begin{equation}\label{f_def}
  \frac{2}{3}\zeta^{3/2}=f(z)\quad, \quad
  f(z)=\ln\left(\frac{1+\sqrt{1-z^2}}{z}\right)-\sqrt{1-z^2}.
\end{equation}
For $z<1$ we can use the behavior of the Airy functions for large
negative argument to simplify the expressions,
\begin{subequations}
\bea
  K_{i\eta}(\eta z) &\sim& \sqrt{\frac{2\pi}{\eta}}
  \frac{e^{-\eta \pi/2}}{(1-z^2)^{1/4}}
  \cos\left(\eta f(z) -\frac{\pi}{4}\right),
\\
  \tilde I_{i\eta}(\eta z) &\sim& -\sqrt{\frac{1}{2\pi\eta}}
  \frac{e^{\eta \pi/2}}{(1-z^2)^{1/4}}
  \sin\left(\eta f(z) -\frac{\pi}{4}\right).
\eea
\end{subequations}
If we choose the branch cut for equation \eqref{f_def} such that
$\zeta$ is a continuous real function of $z$, then for $z>1$ we can
simplify the expressions to read
\begin{subequations}
\bea
  K_{i\eta}(\eta z) &\sim& \sqrt{\frac{\pi}{2\eta}}
  \frac{e^{-\eta \pi/2}}{(z^2-1)^{1/4}}
  e^{-\eta g(z)},
\\
\tilde I_{i\eta}(\eta z) &\sim& \sqrt{\frac{1}{2\pi\eta}}
  \frac{e^{\eta \pi/2}}{(z^2-1)^{1/4}}
  e^{\eta g(z)},
\eea
\end{subequations}
where $g(z)$ is the natural extension of $f(z)$
\begin{equation}
g(z)=-\arcsec z+\sqrt{z^2-1}.
\end{equation}

%%%%%%%%%%%%%%%%%%%%%%%%%%%%%%%%%%%%%%%%%%%%%%%%%%%%%%%%%%%%%%%%%%%

\end{document}